\documentclass[trackchanges]{aastex7}
\usepackage{amsmath}

\usepackage[utf8]{inputenc}
\usepackage{verbatim}
\usepackage{kotex}

\DeclareUnicodeCharacter{039B}{\ensuremath{\Lambda}}

\newcommand{\z}[1]{{\textcolor{red}{#1}}}
\newcommand{\y}[1]{{\textcolor{green}{#1}}}
\usepackage{lineno}

\usepackage{hyperref}
\begin{document}

\title{Warped Disk Galaxies. II. From the Cosmic Web to the Galactic Warp}

\author[orcid=0000-0003-0960-687X, sname='Zee']{Woong-Bae G. Zee}
\affiliation{School of Liberal Studies, Sejong University, Seoul, 05006, Republic of Korea}
\email[show]{galaxy.wb.zi@gmail.com}  

\author[orcid=0000-0001-5512-3735,sname='Jung']{S. Lyla Jung}
\affiliation{Sub-department of Astrophysics, Department of Physics, University of Oxford, Oxford OX1 3RH, UK}
\email[show]{lyla.jung@physics.ox.ac.uk}

\author[orcid=0000-0003-2922-6866,sname='Paudel']{Sanjaya Paudel}
\affiliation{Department of Astronomy, Yonsei University, Seoul, 03722, Republic of Korea}
\affiliation{Center for Galaxy Evolution Research, Yonsei University, Seoul, 03722, Republic of Korea}
\email[show]{sanjpaudel@gmail.com}

\author[orcid=0000-0002-1842-4325, sname='Yoon']{Suk-Jin Yoon}
\affiliation{Department of Astronomy, Yonsei University, Seoul, 03722, Republic of Korea}
\affiliation{Center for Galaxy Evolution Research, Yonsei University, Seoul, 03722, Republic of Korea}
\email[show]{sjyoon0691@yonsei.ac.kr}

\begin{abstract}

Galactic warps are common in disk galaxies. 
While often attributed to galaxy–galaxy tides, a non-spherical dark matter (DM) halo has also been proposed as a driver of disk warping. 
We investigate links among warp morphology, satellite distribution, and large-scale structure using the Sloan Digital Sky Survey catalog of warped disks compiled by \citet{2022ApJ...935...48Z}. 
Warps are classified into 244 S and 127 U types, hosting 1,373 and 740 satellites, respectively, and are compared to an unwarped control matched in stellar mass, redshift, and local density. 
As an indirect, population-level proxy for the host halo’s shape and orientation, we analyze the stacked spatial distribution of satellites.
Warped hosts show a significant anisotropy: an excess at $45^\circ<\phi<90^\circ$ (measured from the host major axis), peaking at $P(\phi)\simeq 0.003$, versus nearly isotropic controls.
Satellites of S type warps preferentially align with the nearest cosmic filament, whereas those of U type warps are more often perpendicular. The incidence of warps increases toward filaments ($r_\textrm{Fila} < 4\,\textrm{Mpc}\,h^{-1}$), while the number of satellites around warped hosts remains approximately constant with filament distance, indicating a direct influence of the large-scale environment.
We discuss possible links between galactic warps and the cosmic web, including anisotropic tidal fields and differences in evolutionary stage.

\end{abstract}


\section{Introduction} \label{sec:intro}

Galactic warps are common in disk galaxies, observed in $\sim60$\,\% of systems (e.g., \citealt{1990MNRAS.246..458S}; \citealt{1998A&A...337....9R}; \citealt{2002A&A...382..513R}; \citealt{2002A&A...391..519C}; \citealt{2002A&A...394..769G}; \citealt{2003A&A...399..457S}; \citealt{2006NewA...11..293A}; \citealt{2016MNRAS.461.4233R}; \citealt{2022ApJ...935...48Z} (hereafter \href{https://iopscience.iop.org/article/10.3847/1538-4357/ac7462/meta}{Paper I})).
Traditionally, warps are classified as S and U type, depending on whether the two sides of the disk bend in opposite directions or the same direction, respectively (\citealt{1998A&A...337....9R}; \citealt{2006NewA...11..293A}; \href{https://iopscience.iop.org/article/10.3847/1538-4357/ac7462/meta}{Paper I}).

Tidal interactions with companion galaxies can readily produce S type warps (e.g., \citealt{1998A&A...337....9R}; \citealt{2001A&A...373..402S}; \citealt{2014ApJ...789...90K}; \citealt{2017MNRAS.465.3446G}; \citealt{2018MNRAS.481..286L}; \citealt{2018Natur.561..360A}).
For example, the Milky Way's prominent warp (\citealt{1957AJ.....62...93K}; \citealt{1988A&A...194..107M}; \citealt{2019NatAs...3..320C}; \citealt{2020ApJS..249...18C}) is attributed to interactions with the Magellanic Clouds and the Sagittarius dwarf spheroidal galaxy (\citealt{2018MNRAS.481..286L}; \citealt{2018Natur.561..360A}). 
The Large Magellanic Cloud (LMC) is the most massive satellite galaxy and likely on its first infall (\citealt{2013ApJ...764..161K}). 
Numerical models of this first-infall show that an LMC-induced wake can warp the disk in a manner consistent with HI constraints and generate a dark matter (DM) halo tracer (\citealt{2018MNRAS.473.1218L}; \citealt{2006ApJ...643..881L}; \citealt{2019ApJ...884...51G}). 
Similarly, the Sagittarius dwarf represents another example of a massive satellite reshaping the Milky Way, undergoing long-term disruption with a higher inferred initial mass, has perturbed the Galaxy over multiple pericenters and can drive spiral structure and bending waves that reproduce solar-neighborhood fluctuations and outer-disk features (\citealt{2011Natur.477..301P}; \citealt{2013MNRAS.429..159G}; \citealt{2017MNRAS.464..794G}; \citealt{2018MNRAS.481..286L}; \citealt{2021MNRAS.508.1459H}).
While these interaction models generally reproduce the observed phase and location of the Galactic warp, they tend to underpredict its amplitude (\citealt{2015ApJ...802..128G}).
This limitation reflects that many early models assumed a static Galactic potential and ignored the reflex response to a massive satellite. More recent $N$-body simulations include this effect and demonstrate its role in amplifying disk perturbations (e.g., \citealt{2019MNRAS.487.2685E}; \citealt{2020ApJ...893..121P}; \citealt{2021ApJ...923..149S}). 
However, broader cosmological simulations reveal that satellite interactions alone cannot account for the ubiquity of warp morphologies.
Such simulations struggle to explain U type warps.
Using the IllustrisTNG simulations, \citet{2020MNRAS.498.3535S} identified 249 S type warped galaxies, finding that $\sim30$\,\% are linked to recent tidal interactions, while U type warps are absent, suggesting that tidal mechanisms alone may be insufficient to account for all warp origins.

To address limitations of existing warp formation models, alternative mechanisms have been proposed, including galactic magnetic field (\citealt{1990A&A...236....1B}; \citealt{1998A&A...332..809B}), cosmic infall (\citealt{1989MNRAS.237..785O}; \citealt{1999MNRAS.303L...7J}; \citealt{2010MNRAS.408..783R}), ram-pressure stripping (RPS) (\citealt{2014MNRAS.440L..21H}; \href{https://iopscience.iop.org/article/10.3847/1538-4357/ac7462/meta}{Paper I}), and misaligned DM halos (\citealt{1988MNRAS.234..873S}; \citealt{1995MNRAS.275..897N}; \citealt{2005ApJ...627..647B}; \citealt{2005ApJ...627L..17B}; \citealt{2009ApJ...696.1899J}; \citealt{2009ApJ...703.2068D}; \citealt{2022MNRAS.510.1375S}; \citealt{2023MNRAS.523.5853V}). 
In \href{https://iopscience.iop.org/article/10.3847/1538-4357/ac7462/meta}{Paper I}, the first study in the \textit{Warped Disk Galaxies} series, an analysis of 2,206 S type and 1,456 U type warps in Sloan Digital Sky Survey (SDSS) Data Release 7 (DR7) showed that, at fixed stellar mass, U types are bluer and of higher sSFR than S types; S types track the tidal force of the nearest neighbor; by contrast, U types—especially in clusters—show no clear tidal correlation, indicating a distinct origin.
U types in clusters are more concentrated into the cluster's center and have lower relative velocities than S types, consistent with earlier cluster infall and enhanced RPS, in line with \citet{2012MNRAS.420.1990S} who demonstrated that RPS can displace a galaxy’s DM halo and bend the stellar disk into a U-shaped warp.
\href{https://iopscience.iop.org/article/10.3847/1538-4357/ac7462/meta}{Paper I} estimated that at least $17.5\%$ of U type warps in clusters can be explained by RPS; however, still the majority of U type warps in the sample lie outside clusters, underscoring the need for a more general mechanism that can operate across environments.

Building on this foundation, this study investigates whether galactic warps are connected to DM halo structure in a cosmological context.
Early work showed that halo properties can regulate warp formation even in isolation: \citet{1988MNRAS.234..873S} identified core radius and flattening as key parameters, with oblate halos supporting evolving warps without external perturbers.
Subsequent studies established that halos are typically triaxial and become more spherical with radius, naturally introducing radial twists and disk–halo misalignment whose torques can excite and sustain warps over long timescales; warps in oblate halos are generally weaker than those from strong tidal interactions (\citealt{2005ApJ...627..647B}; \citealt{2005ApJ...627L..17B}).
Simulations further indicated that most lopsided/warped disks inhabit asymmetric halos, with limited evidence for a purely tidal origin (\citealt{2023MNRAS.523.5853V}).
Observational evidence is consistent with this picture: in the Virgo Cluster, warp strength decreases with increasing dynamical mass (\citealt{2012AAS...21934614B}); in the Milky Way, stellar age gradients and retrograde disk precession imply a slightly oblate, tilted halo relative to the stellar disk (\citealt{2023NatAs...7.1481H}; \citealt{2024NatAs.tmp..189H}).
Most recently, zoomed hydrodynamical simulations by \citet{2025ApJ...988..190A} found a persistent, filament-aligned quadrupole in the zoomed cosmological-baryonic simulations of Milky Way-mass halos.
Specifically, an LMC-mass satellite excites a quadrupolar response that is an order of magnitude stronger in such a realistic, triaxial, time-evolving halo than in isolated spherical halo setups.

Cosmic accretion, via gradual infall from the intergalactic medium, can torque a galaxy’s DM halo and embedded disk, reorienting their angular momenta and inducing warps (\citealt{1989MNRAS.237..785O}; \citealt{1999MNRAS.303L...7J}). 
Within the framework of tidal torque theory (TTT; \citealt{1969ApJ...155..393P,1970Afz.....6..581D}), large-scale anisotropic inflows dominate the buildup of galaxy spin and naturally produce misaligned, warped disks. 
\citet{2008A&ARv..15..189S} further suggested that the ubiquity of warped disks reflects ongoing cold-gas accretion from the cosmic web. 
High-resolution cosmological simulations by \citet{2015MNRAS.449.2087D} demonstrated that most galactic angular momentum at $1.5 < z < 4$ is acquired through cold flows along large-scale filaments. 
\citet{2017MNRAS.466.4692K} showed that galaxies near cosmic filaments exhibit elevated HI gas fractions, suggesting that filamentary accretion builds larger gas reservoirs that make disks more susceptible to warping (see also \citealt{2002A&A...394..769G}; \citealt{2003A&A...399..457S}; \citealt{2022MNRAS.513.2168T}). 
The prevalence of warps, even out to $z \sim 2$ (e.g., \citealt{2025A&A...697L...1R}), may imply a consequence of early cosmic inflows. 
Collectively, these results place warp formation in a cosmological context governed by large-scale accretion.

Recent advances in extensive galaxy surveys and cosmological mapping now enable systematic studies of large-scale environmental impacts on galactic morphological evolution. 
Surveys such as the SDSS (\citealt{2000AJ....120.1579Y}; \citealt{2020ApJS..249....3A}; \citealt{2014MNRAS.438.3465T}), the Galaxy and Mass Assembly survey (GAMA; \citealt{2011MNRAS.413..971D}; \citealt{2014MNRAS.438..177A}), the Dark Energy Spectroscopic Instrument (DESI; \citealt{2019AJ....157..168D}; \citealt{2024MNRAS.534.3540L}), and the Euclid Quick Data Release (\!\citealt{2025arXiv250315333E}) have provided unprecedented coverage for reconstructing the cosmic web. 
Filament identification techniques such as the T-Web (\citealt{2024A&A...686A.276A}), V-Web (\citealt{2017ApJ...845...55P}), and the DIScrete PERsistent Structures Extractor (DisPerSE; \citealt{2011MNRAS.414..350S}; \citealt{2011MNRAS.414..384S}) allow robust tracing of filamentary structures in both simulations and observations. 
These tools have revealed that large-scale structure strongly influences galaxy properties. 
Galaxies in filaments differ from those in voids, being more often early-type and quenched (\citealt{2016MNRAS.457.2287A}; \citealt{2017A&A...597A..86P}; \citealt{2017A&A...600L...6K}; \citealt{2018MNRAS.474.5437L}; \citealt{2020MNRAS.491.4294K}; \citealt{2020A&A...639A..71K}; \citealt{2024MNRAS.528.4139H}; \!\citealt{2025arXiv250315333E}).
Moreover, \citet{2022MNRAS.510.3071R} found that galaxy groups in filaments are more elongated in shape and show no strong correlation between star formation and local density, suggesting that cosmic web effects can overshadow local environmental influences.
Such findings highlight the critical role of the cosmic web, beyond local interactions, in shaping galaxy evolution.

As expected, the cosmic web imprints itself on DM halos.
Cosmological simulations showed that halo spins and shapes tend to align with nearby filaments, with low-mass halos retaining spins along filaments while high-mass halos often flip to perpendicular orientations as they grow (\citealt{2006MNRAS.370.1422A}; \citealt{2007ApJ...655L...5A}; \citealt{2012MNRAS.427.3320C}; \citealt{2013MNRAS.428.2489L}; \citealt{2015MNRAS.446.2744L}; \citealt{2017MNRAS.468L.123W}; \citealt{2018MNRAS.481..414G}; \citealt{2019MNRAS.487.1607G}; \citealt{2025MNRAS.539.1692Z}).
Halo spin--filament alignments evolve over cosmic time (e.g., \citealt{2013ApJ...762...72T}; \citealt{2018MNRAS.481.4753C}), and coherent spin fields aligned with filaments can drive correlated galaxy rotations or seed intergalactic magnetic fields aligned with filament axes (\citealt{2021MNRAS.506.1059X}; \citealt{2021MNRAS.503.4016B}). 
Given that changes in DM halo angular momentum have been proposed as a plausible driver of galactic warps (independent of direct tidal perturbations), these results imply that the cosmic web could influence the formation of warps primarily through its impact on the dynamics of DM halos.

Simulations by \citet{1988MNRAS.234..873S} first suggested that the core radius of DM halos is a critical parameter influencing galactic warp formation, showing that isolated disks embedded in oblate halos can develop warps without external interactions. 
\citet{2005ApJ...627..647B} and \citet{2005ApJ...627L..17B} demonstrated that DM halos typically exhibit triaxial shapes, and that misalignments between their inner and outer regions can induce tilted DM infall, generating torques on stellar disks that excite and sustain stable warps. 
More recently, \citet{2023MNRAS.523.5853V} used IllustrisTNG simulations to show that the vast majority of lopsided, warped disk galaxies reside within asymmetric DM halos, with little evidence for tidal origin. 
Observational evidence, while limited, also pointed to DM halo's role: more massive (halo-dominated) galaxies tend to have weaker warps in the Virgo Cluster (\citealt{2012AAS...21934614B}), and in the Milky Way, the measured precession of the warp implies its DM halo is slightly tilted relative to the disk (\citealt{2023NatAs...7.1481H}; \citealt{2024NatAs.tmp..189H}).

The shape of DM halos can be probed by the spatial distribution of satellite galaxies (e.g., \citealt{1995MNRAS.275..429L}; \citealt{1997ApJ...478L..53Z}; \citealt{2011MNRAS.415.2607D}; \citealt{2014ApJ...791L..33D}; \citealt{2019MNRAS.489..459B}; \citealt{2021ApJ...914...78W}; \citealt{2023ApJ...947...56S}; \citealt{2024MNRAS.529.1405L}).
Over the past few decades, numerous studies have reported lopsided satellite distributions (LSDs) as observational evidence for the anisotropic nature of DM halos (e.g., \citealt{2005ApJ...629..219Z}; \citealt{2008MNRAS.390.1133B}; \citealt{2016ApJ...830..121L}; \citealt{2017ApJ...850..132P}; \citealt{2019MNRAS.488.3100G}; \citealt{2020ApJ...898L..15B}; \citealt{2021ApJ...914...78W}; \citealt{2023ApJ...947...56S}). 
For instance, \citet{2008MNRAS.385.1511W} analyzed SDSS DR4 galaxy groups and found that massive groups preferentially host prolate halos, whereas low-mass groups tend to have nearly spherical halos, suggesting a mass-dependent halo evolution. 
Even isolated disk galaxies from SDSS DR11 exhibit anisotropic satellite arrangements (\citealt{2020ApJ...898L..15B}), underscoring that DM halos are generally not spherical. 
Following these findings, the spatial distribution of satellites around warped galaxies may serve as an effective tracer of DM halo shapes, which stem from the cosmic web and, in turn, influence warp formation. 
This provides a potential observational pathway to probe the connection between galactic warps and large-scale filaments, with DM halos acting as the intermediate link.

In this paper, we investigate the cosmological origin of galactic warps by using the spatial distribution of satellites as a proxy for DM halo shapes, which are expected to be shaped by large-scale filaments. 
This paper is organized as follows. 
Section~\ref{sec:data} describes the warped galaxy sample and the unwarped control sample. 
Section~\ref{sec:result} presents our analysis of warp frequency and amplitude as a function of distance to the nearest filament, as well as the preferred orientations of satellites around warped galaxies, suggesting anisotropic halo influences. 
We further show that these anisotropic DM halos are aligned with nearby filaments, with the alignment signal varying by warp morphology. 
Section~\ref{sec:web} discusses the role of large-scale cosmic webs in shaping DM halos and driving warp formation, and Section~\ref{sec:sum} summarizes our main results. 
Throughout this paper, we adopt a flat $\Lambda$CDM cosmology with $\Omega_\textrm{m} = 0.3$, $\Omega_\Lambda = 0.7$, $H_0 = 70~\textrm{km s}^{-1}\textrm{Mpc}^{-1}$ (i.e., $h=0.7$).

\section{Data and Methodology} \label{sec:data}

\subsection{Galactic Warps}

Our sample selection and warp identification methodology closely follows the approach outlined in \href{https://iopscience.iop.org/article/10.3847/1538-4357/ac7462/meta}{Paper I}. 
The galaxies analyzed in this study are drawn from the SDSS DR7 (\citealt{2009ApJS..182..543A}) using CasJob queries. 
Since warped disk structures can only be identified in edge-on galaxies, we selected highly inclined edge-on disk galaxies with $M_*$\,$>$\,$10^9M_\odot$ in the local universe ($0.02 < z < 0.06$) from the Main Galaxy Sample ({\fontfamily{qcr}\selectfont PhotoObjAll.type = 3} and {\fontfamily{qcr}\selectfont SpecObjAll.specClass = 2}). 
To ensure our sample consists of genuine edge-on disk galaxies, we applied the morphological classifications from the Galaxy Zoo 2 (GZ2) project (\citealt{2013MNRAS.435.2835W}), selecting galaxies classified as edge-on spirals by volunteers with an edge-on probability greater than $0.5$. 
Additionally, to confirm a high inclination, we use photometric bulge+disk decomposition data (\citealt{2011ApJS..196...11S}) and excluded galaxies with inclinations smaller than 60$^\circ$ ($i < 60^\circ$).

\href{https://iopscience.iop.org/article/10.3847/1538-4357/ac7462/meta}{Paper I} developed an automated method for detecting and quantifying galactic warps, which we briefly summarize here (see also Figure 1 in \href{https://iopscience.iop.org/article/10.3847/1538-4357/ac7462/meta}{Paper I}).
Using $500 \times 500$ pixel ($3.3' \times 3.3'$) FITS images centered on each target galaxy, the procedure involves the following steps: 
\begin{enumerate}
    \item Identify the galaxy's major axis. 
    Re-estimate the major axis and re-rotate iteratively until the position angle (PA) meets a strict tolerance ($|{\rm PA}|<0.01^\circ$).
    \item Trace the \textit{`spine'} by connecting the brightest points along the Gaussian-fitted vertical brightness profile at each x-coordinate across the disk.
    \item Measure the warp amplitude as the degree of misalignment of the central major axis relative to the outermost warped edges on both sides. 
    \item Classify the warp morphology into two main types: 
    \begin{itemize}
    \item S type: Both disk ends bend in opposite directions.
    \item U type: Both ends bend in the same direction. 
    \end{itemize}
\end{enumerate}

This approach identified 2206 (27.6\%) S type warps and 1456 (18.2\%) U type warps among $\sim$8000 edge-on disk galaxies (see also Figure 2 and Table 1 in \href{https://iopscience.iop.org/article/10.3847/1538-4357/ac7462/meta}{Paper I}).
The overall warp incidence of $\sim$46\%, along with the finding that S type warps are twice as common as U types, is consistent with previous observations, which have shown that galactic warps are a common structural feature (\citealt{1998A&A...337....9R}; \citealt{2002A&A...382..513R}; \citealt{2003A&A...399..457S}; \citealt{2006NewA...11..293A}; \citealt{2016MNRAS.461.4233R}).
Building on this analysis, we also launched the \textit{Poppin' Galaxy}\footnote{\url{https://www.zooniverse.org/projects/wim0705/poppin-galaxy}} project on Zooniverse to collect extensive morphological classifications of warped disk galaxies. 
The project has yielded over $\sim11,000$ classifications from more than $\sim5,000$ volunteers, providing a valuable training set for a forthcoming deep learning-based framework for automated warp identification.

\subsection{Large-scale Filaments}

To investigate the cosmological origins of galactic warps, we utilize the cosmic filament catalog from \citet{2020A&A...642A..19M}, constructed using DisPerSE---a widely used open-source algorithm for identifying the topology of the cosmic web. 
DisPerSE has been extensively applied in both observational (e.g., \citealt{2020A&A...642A..19M}; \citealt{2021MNRAS.504.4626K}; \citealt{2021MNRAS.505.4920W}; \citealt{2025MNRAS.539.2362J}) and theoretical works (e.g., \citealt{2020MNRAS.493..362K}; \citealt{2020A&A...641A.173G}; \citealt{2024A&A...684A..63G}; \citealt{2025MNRAS.538..830W}). 
We use the filament catalogue based on the SDSS DR12 LOWZ+CMASS sample (\citealt{2015ApJS..219...12A}; \citealt{2016MNRAS.455.1553R}) with one round of density field smoothing, a persistence threshold of $3\sigma$, and one skeleton smoothing pass. 
These settings mitigate noise and ensure that only physically significant filaments are retained.

The \citet{2020A&A...642A..19M} catalog\footnote{\url{https://l3s.osups.universite-paris-saclay.fr/cosfil.html}} provides 3D coordinates (RA, Dec., redshift) of filament sampling points. 
For each galaxy among our sample, we identify the nearest filament point using \texttt{astropy.coordinates.match\_coordinates\_3d} function, and compute the distance to the filament (see also Figures 3 and 5 of \citealt{2025MNRAS.539.2362J}, which illustrate filament-orientation measurements in Cartesian and spherical coordinates). 
The local orientation of the nearest filament point is determined the same way as \citet{2025MNRAS.539.2362J}; we define the filament orientation as the vector connecting two adjacent sampling points on the same filament, using the nearest point and its neighbor when the nearest point lies at a filament endpoint, following \citet{2007ApJ...671.1248L}. 
We compute the filament position angle as: \begin{equation} PA_\textrm{Fila} = \textrm{arctan}\left(\cfrac{-\nu_{\phi}}{\nu_{\theta} \cos(\textrm{Dec.})}\right), \end{equation} where $(\nu_r, \nu_\theta, \nu_\phi)$ are the radial, polar, and azimuthal components of the filament orientation vector in the local spherical frame.

\subsection{Spatial Distribution of Satellite Galaxies}

To probe the shape and orientation of DM halos, stacked satellite anisotropy around central galaxies serves as a proxy for halo geometry and thus reflects population-level trends; given the few satellites per host, any inference of individual-halo principal axes requires particular caution.
Isolated central galaxies are drawn from the warped-disk sample of \href{https://iopscience.iop.org/article/10.3847/1538-4357/ac7462/meta}{Paper I}; satellites are identified by projected separation and line-of-sight velocity differences following \citet{2020ApJ...898L..15B} and \citet{2023ApJ...947...56S}.
Central galaxies come from the \citet{2014ApJS..210....3M} stellar-mass catalog with $M_\ast>10^9\,M_\odot$ and are required to be at least 1 mag brighter (in $g$ band) than any neighbor within $r_p<750\,\mathrm{kpc}\,h^{-1}$ and $|\Delta v|<1000\,\mathrm{km\,s^{-1}}$. 
Satellites satisfy $r_p<500\,\mathrm{kpc}\,h^{-1}$, $|\Delta v|<500\,\mathrm{km\,s^{-1}}$, and $M_\ast>10^7\,M_\odot$.
All centrals are edge-on disk galaxies suitable for warp analysis, but no morphological or inclination constraints are applied to satellites. 
Edge effects are minimized by excluding centrals within $4\,\mathrm{Mpc}\,h^{-1}$ of the survey boundary. 
To ensure minimal angular sampling, each central galaxy must host $\geq3$ satellites \citep{2008MNRAS.385.1511W}. 
The final sample comprises 244 S type and 127 U type centrals hosting 1,373 and 740 satellites, respectively (mean $\sim$ 5–6 satellites per host), underscoring that results are interpreted as ensemble anisotropies rather than per-halo shapes.
Although this reduced the available sample, an additional analysis restricted to hosts with $\geq 6$ satellites was employed to improve the reliability of host-level inferences of DM halo geometry. 
This subset comprises 95 S type centrals (654 satellites in total) and 58 U type centrals (408 satellites in total).

\subsection{Control Sample}

To assess differences in satellite distribution between warped and unwarped galaxies, we construct a meticulously matched control sample, applying identical satellite selection criteria. 
To minimize bias from morphology and inclination, we select highly inclined, edge-on disk galaxies without detectable warps. 
Control centrals are also required to satisfy the central galaxy criteria, hosting at least three satellites within $r_p$\,$<$\,$500$\,$\textrm{kpc}$\,$h^{-1}$ and $|\Delta v|$\,$<$\,$500$\,$\textrm{km}$\,$\textrm{s}^{-1}$. 
To remove intrinsic biases, control galaxies are matched to warped galaxies in redshift, stellar mass, and local density. 
The local density, $\textrm{Log}\,\Sigma$, is estimated using the adaptive kernal method (\citealt{1986desd.book.....S}), following the approach of previous works (\citealt{2011A&A...531A.114F}; \citealt{2015ApJ...805..121D}; \citealt{2019ApJ...882...14M}; \citealt{2024ApJ...963..141Z}).
For each warped galaxy, one control is randomly selected within $\pm0.005$ in redshift, $\pm0.01$ dex in stellar mass, and $\pm0.01$ dex in local density, and this procedure is randomly repeated to ensure statistical robustness. 
The final control sample consists of 4,347 unwarped edge-on centrals and 14,512 associated satellites.

\section{Result}
\label{sec:result}

\subsection{Anisotropic Distribution of Satellites around Warped Disk Galaxies}
\label{sec:excess}

\begin{figure}
	\includegraphics[width=\columnwidth]{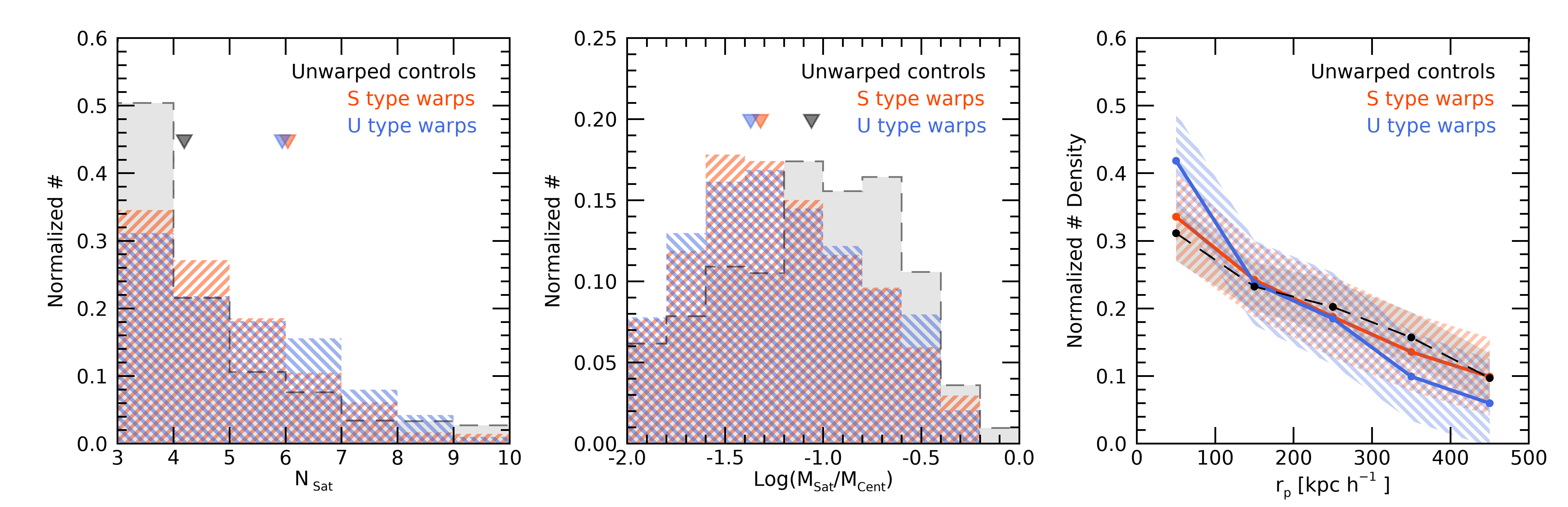}
    \caption{Left: The normalized histogram of the number of satellites per central galaxy for the unwarped control sample (dashed black), S type (red), and U type warped galaxies (blue).
    The mean satellite count for each category is described by triangle symbols, showing that warped galaxies systematically host more satellites than their unwarped counterparts.
    Middle: The distribution of satellite-to-central stellar mass ratios for the same three groups. 
    The mean mass ratio is displayed as triangle symbols, revealing that warped galaxies tend to host lower-mass satellites on average.
    Right: The radial number density of satellites as a function of projected separation ($r_{\rm p}$) from the central galaxy, for unwarped control galaxies (dashed black), S type (red), and U type (blue) warped galaxies. 
    Shaded region represents Poisson errors for each bin.}
    \label{fig:1}
\end{figure}

Using the warp identification and measurement method of \citet{2024ApJ...963..141Z}, we find that warped galaxies host a significantly greater number of satellite galaxies compared to unwarped controls. 
The left panel of Figure~\ref{fig:1} shows that both S and U type warped galaxies tend to host more satellites, with S types averaging $6.04\,\pm\,3.28$ satellites (range 3--25) and U type averaging $5.94\,\pm\,2.85$ (range 3--21), whereas unwarped galaxies host satellites with $4.19\,\pm\,1.69$ (range 3--12).
Although the mean and median values of each sample agree within $1\sigma$, a two-sample Kolmogorov–Smirnov (KS) test shows that the distributions of satellite numbers differ from the control sample (control vs. S type: $D=0.179$, $p=4.6\times10^{-14}$; control vs. U type: $D=0.231$, $p=1.24\times10^{-10}$).
The middle panel demonstrates that satellites of warped galaxies have lower mass ratios relative to their central galaxies. 
Unwarped systems show a mean mass ratio of $\langle \log (M_\textrm{Sat}/M_\textrm{Cent}) \rangle = -1.06 \pm 0.44$, while S and U type warped galaxies exhibit lower values of $-1.32 \pm 0.45$ and $-1.37 \pm 0.48$, respectively. 
Consistently, two-sample KS tests show significant shifts in the mass ratio distributions relative to controls (control vs. S type: $D=0.233$, $p=1.7\times10^{-21}$; control vs. U type: $D=0.181$, $p=7.3\times10^{-10}$).
These results suggest that warped disks tend to host more numerous, relatively low-mass satellites, implying that frequent but weaker interactions may contribute to warp formation or maintenance. 
The right panel shows the normalized number density of satellites as a function of projected separation, $r_{\rm p}$.
For each 100 kpc annulus, we compute the fraction of satellites and normalize by the annulus area to account for coverage differences. 
S type warped galaxies exhibit radial profiles similar to unwarped systems, while U type warped galaxies are more centrally concentrated, with normalized densities $\sim0.4$ higher within $r_{\rm p} < 100\ \textrm{kpc} \ h^{-1}$. 
However, this difference is not statistically significant. 
Thus, proximity alone does not fully reveal the spatial distribution differences among warp morphologies.

\begin{figure}
	\includegraphics[width=\columnwidth]{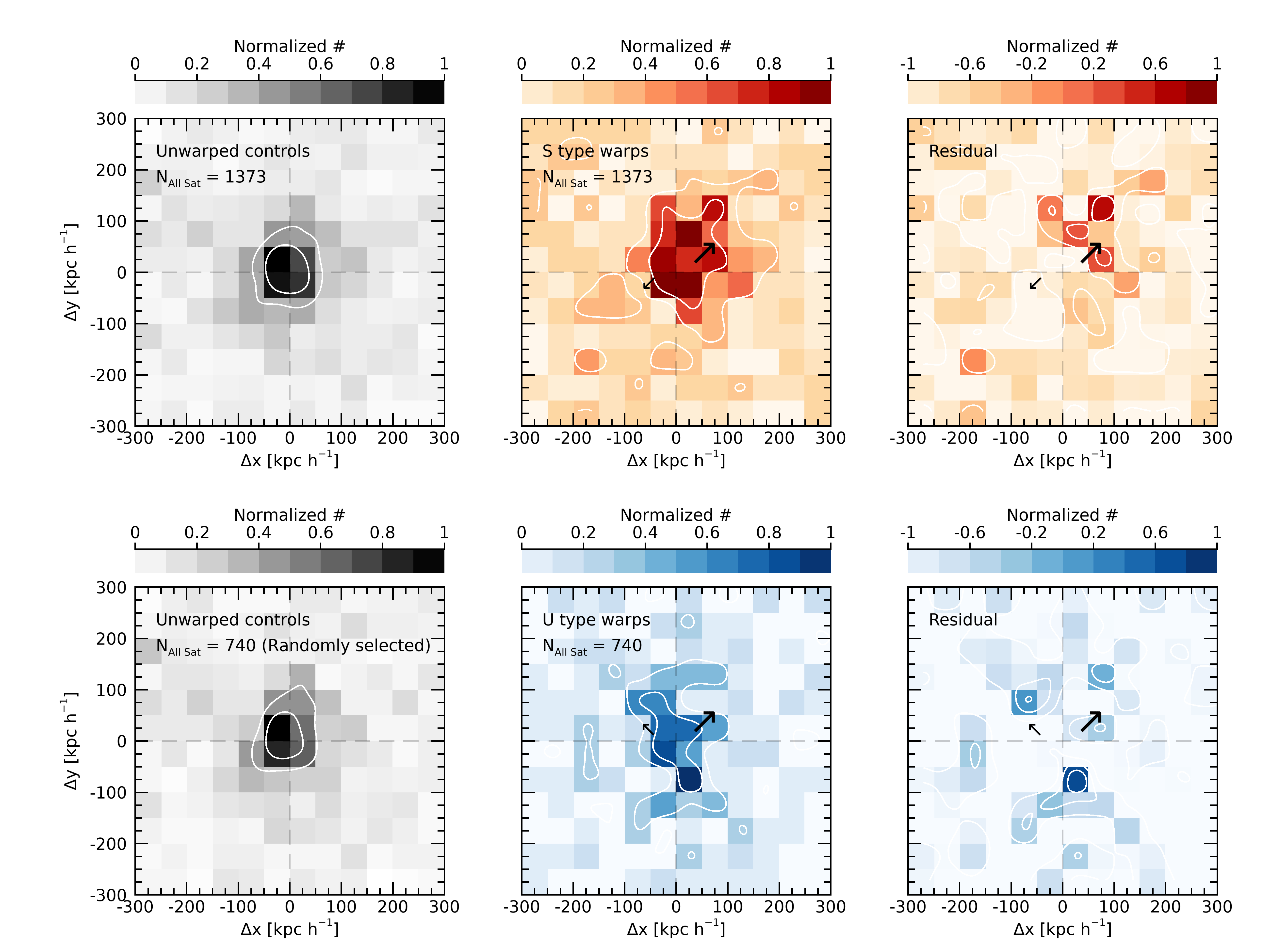}
    \caption{Upper: Density distribution of stacked satellite galaxies in $\Delta x$–$\Delta y$ space relative to their central galaxies for the unwarped control sample (left), S-type warped galaxies (middle), and the residual map obtained by subtracting the control sample from the S-type distribution (right). 
    White contours are drawn at 0.5$\sigma$ intervals. 
    An equal number of all stacked satellites, randomly selected from the control sample, is used as indicated. 
    Larger and smaller black arrows mark the directions of the stronger and weaker warp sides, respectively.
    Lower: Same as the upper panels, but for U-type warped galaxies.}
    \label{fig:2}
\end{figure}

\begin{figure}
	\includegraphics[width=\columnwidth]{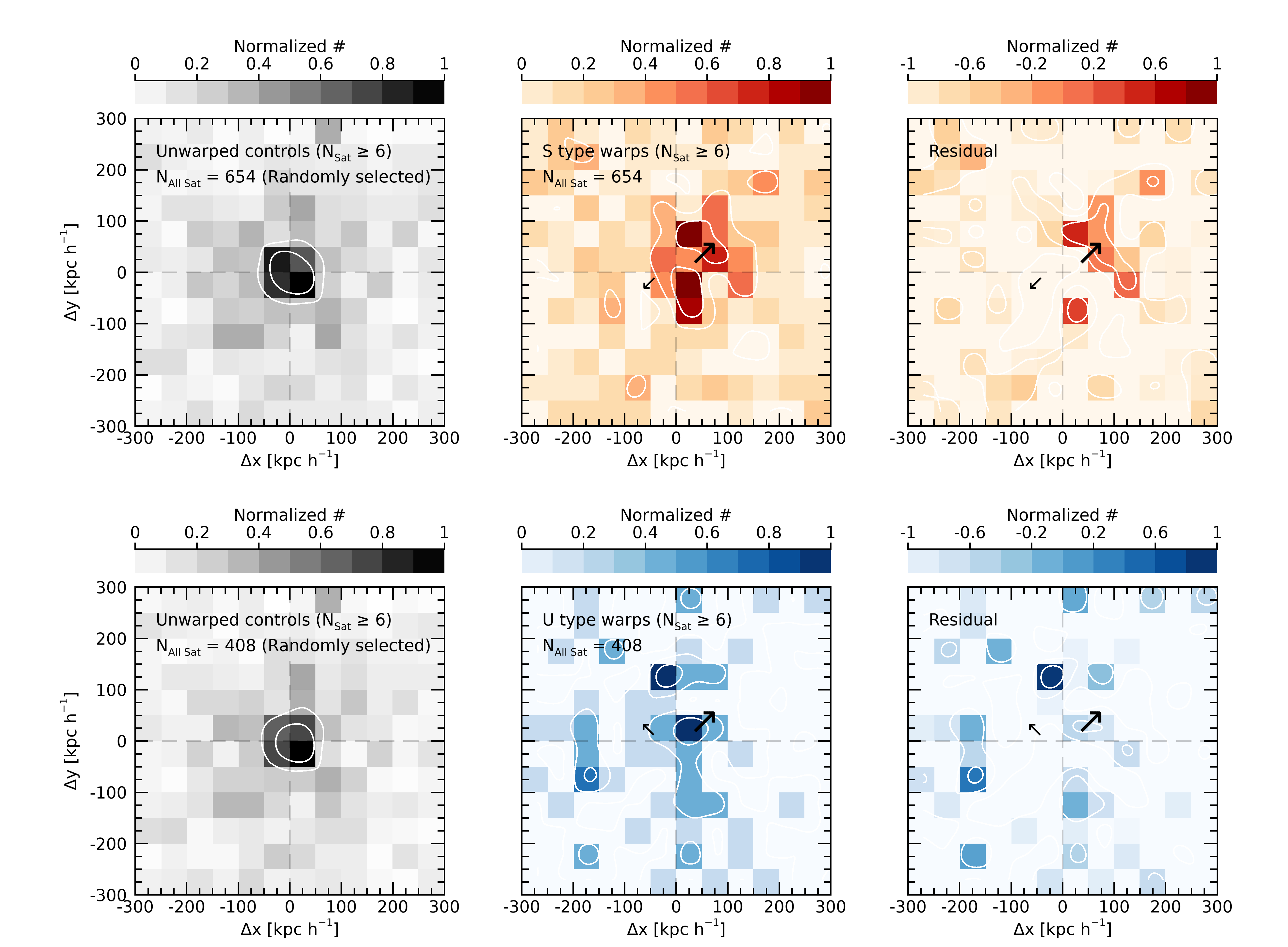}
    \caption{Same as Figure~\ref{fig:2}, but for central galaxies that host at least six satellites ($N_\textrm{Sat}\ge 6$).}
    \label{fig:2b}
\end{figure}

Figure~\ref{fig:2} presents the stacked satellite density distributions in the $\Delta x$–$\Delta y$ plane for S and U type wars, respectively.
$\Delta x$ and $\Delta y$ are the projected separations along the major and minor axes of the central galaxy, respectively. 
To analyze their spatial structure, we align each galaxy's major axes horizontally. 
For warped galaxies, we define the strongly warped side as the first quadrant and reorient the images accordingly; weaker sides for S types appear in the third quadrant (upper row), and for U types in the second quadrant (lower row). 
Large and smaller arrows indicate strong and weaker warp sides. 
For unwarped controls, lacking defined warp directions, we mirror galaxies randomly along vertical or horizontal axes to ensure fair comparisons. 
Additionally, in Figure~\ref{fig:2}, to match satellite numbers with warped samples, we randomly subsample unwarped satellites: 1373 for S types and 740 for U types, from a total of 14,512. 
Interestingly, satellites around S and U type warped galaxies exhibit more elongated and anisotropic distributions compared to unwarped galaxies differently. 
S types show a clear concentration in the first quadrant, aligning with their stronger warp side (upper row), while U types display elongation along the y-axis, nearly perpendicular to the disk's major axis (lower row).
As illustrated in the right panels of Figure~\ref{fig:2}, residual maps highlight these anisotropies more clearly. 
These findings imply that warped disk galaxies are embedded in more elongated, non-spherical DM potentials, as traced by their satellites' anisotropic distributions, whereas unwarped galaxies host satellites in more isotropic, spherical configurations. 
Figure~\ref{fig:2b} repeats the same analysis for a more restrictive subsample—central galaxies hosting at least six satellites ($N_\textrm{Sat}\ge 6$)—and the trends persist.
Compared to Figure~\ref{fig:2}, the residual distribution shows a modest shift under the $N_\textrm{Sat}\ge 6$ restriction, most noticeably for U type warps. 
To make this explicit, we quantify the spatial anisotropy of satellites around warped and unwarped galaxies as follows.

\begin{figure}
	\includegraphics[width=\columnwidth]{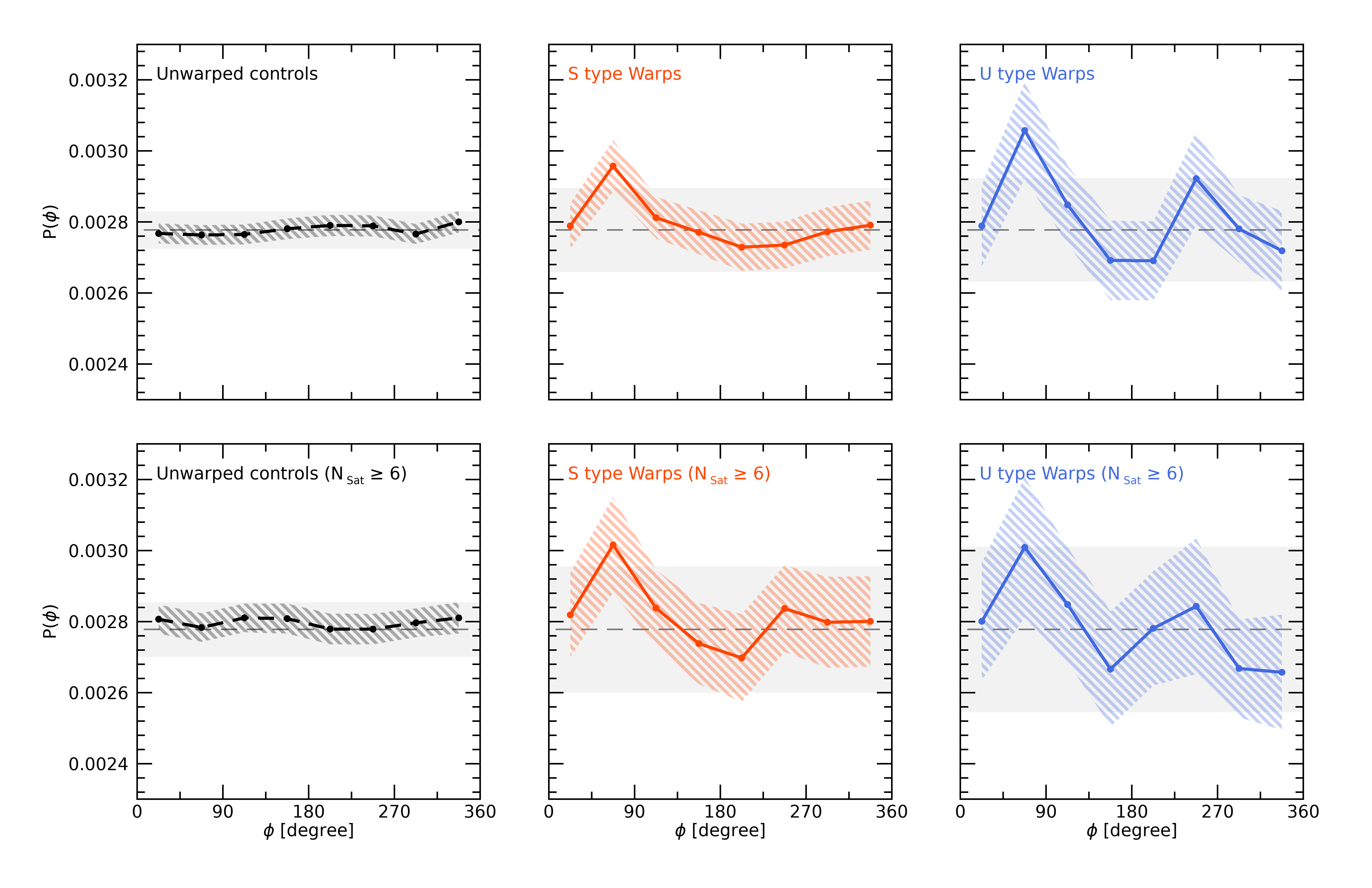}
    \caption{Upper: Probability density function, $P(\phi)$, of satellite position angle, $\phi$, relative to the major axis of the central galaxy, for unwarped control sample (left, black), S type (middle, red), and U type (right, blue) warped galaxies. 
    Shaded region represents Poisson errors for each bin.
    For comparison, the expected isotropic distribution with the same number of central and satellite galaxies is shown as a gray horizontal line, while the 3$\sigma$ range of its random distribution is shaded in gray. 
    For S and U type warped galaxies, significant anisotropies appear, with satellites preferentially clustering at $45^{\circ} < \phi < 90^{\circ}$. 
    Lower: Same as the upper panels, but for central galaxies that host at least six satellites ($N_\textrm{Sat}\ge 6$).}
    \label{fig:2c}
\end{figure}

We define the position angle of each satellite, $\phi$, measured relative to the major axis of the central galaxy. 
In previous studies, satellite distributions were typically examined in terms of whether satellites align preferentially along the major or minor axis of their central galaxy (\citealt{2004MNRAS.348.1236S}; \citealt{2005MNRAS.363..146L}; \citealt{2008ApJ...680..287M}; \citealt{2010ApJ...709.1321A}; \citealt{2012MNRAS.423.1109P}; \citealt{2014ApJ...793L..42B}; \citealt{2015MNRAS.450.2727T}; \citealt{2015MNRAS.452.1052L}; \citealt{2017ApJ...850..132P}; \citealt{2019MNRAS.484.4325W}; \citealt{2019MNRAS.488.3100G}; \citealt{2023ApJ...947...56S}; \citealt{2024AAS...24345911M}). 
These studies generally defined $\phi$ within a range of $0^{^\circ}$ to $90^{^\circ}$, where $\phi = 0^{^\circ}$ corresponds to satellites aligned with the central's major axis, and $\phi = 90^{^\circ}$ corresponds to satellites along the minor axis. 
However, in this study, a more detailed 2D approach is required to examine the relationship between satellite positions and the direction of the warped structures. 
We therefore defined $\phi$ as the misalignment angle between the satellite position and the right-hand side of the major axis of its central galaxy, measured in an anticlockwise direction. 
With this definition, $\phi$ ranges from $0^{^\circ}$ to $360^{^\circ}$, covering all four quadrants in the $\Delta \textrm{x}$--$\Delta \textrm{y}$ space. 
This approach allows for a more precise analysis of how satellite positions correlate with the observed warp morphology, as illustrated in Figure~\ref{fig:2c}.

To statistically estimate the distribution of $\phi$ of satellites, we define the probability density function of $\phi$, denoted as $P(\phi)$. 
This probability is obtained by counting the number of satellites within a given angular bin of $\phi$ and normalizing it by the total number of satellites in the sample. 
Mathematically, it is expressed as \begin{equation} P(\phi) = \cfrac{\textrm{N}(\phi)} {\textrm{N}_\textrm{Total}} \cdot \cfrac{1}{\Delta \phi} \hspace{1mm}, \end{equation} where $\textrm{N}(\phi)$ is the number of satellites within the angular bin centered at $\phi$, $\textrm{N}_\textrm{Total}$ is the total number of satellites in the sample, and $\Delta \phi$ is the bin width used for the histogram (\citealt{2019MNRAS.488.3100G}; \citealt{2023ApJ...947...56S}; \citealt{2024AAS...24345911M}). 
This definition ensures that $P(\phi)$ represents a properly normalized probability density, allowing for a direct comparison of satellite anisotropies between warped and unwarped galaxies.

The upper panels of Figure~\ref{fig:2c} show $P(\phi)$ a function of $\phi$ for all unwarped (left), S type (middle), and U type warped galaxies (right), with gray bands marking the $3\sigma_\textrm{Random}$ range expected for a uniform distribution given the same numbers of centrals and satellites.
For unwarped hosts, $P(\phi)$ is nearly flat ($\sigma \approx 1.42 \times 10^{-5}$) and remains within the band ($3\sigma_\textrm{Random} \approx 5.31 \times 10^{-5}$), consistent with isotropic distribution. 
By contrast, S and U type systems show an anisotrophy: Satellites around S type peak at $45^{\circ} < \phi < 90^{\circ}$ with $P(\phi) \approx 0.0029$, aligning with the stronger warped side of disk; those of U type exhibit a peak in the same range with $P(\phi) > 0.003$, exceeding the uniform thresholds (S type: $3\sigma_\textrm{Random} \approx 7.41 \times 10^{-4}$; U type: $3\sigma_\textrm{Random} \approx 1.18 \times 10^{-4}$). 
As shown in the lower panels, restricting to hosts with $N_\textrm{Sat} \geq 6$ yields qualitatively similar trends, albeit with larger uncertainties owing to the reduced sample size.

Despite modest numbers, the stacked satellite distribution exhibits a pronounced excess aligned with the disk’s stronger-warped side, consistent with scenarios in which satellite tides torque disks and generate warps (\citealt{1998A&A...337....9R}; \citealt{2001A&A...373..402S}; \citealt{2014ApJ...789...90K}; \citealt{2017MNRAS.465.3446G}; \citealt{2018MNRAS.481..286L}; \citealt{2018Natur.561..360A}). 
However, without full 3D satellite kinematics, this anisotropy cannot be attributed solely to recent encounters; similar angular patterns are expected if satellites trace a non-spherical dark-matter halo (e.g., triaxial or filament-aligned quadrupoles; \citealt{1988MNRAS.234..873S}; \citealt{2005ApJ...627..647B}; \citealt{2005ApJ...627L..17B}; \citealt{2023MNRAS.523.5853V}; \citealt{2025ApJ...988..190A}; \citealt{2023NatAs...7.1481H}; \citealt{2024NatAs.tmp..189H}). 
Taken together, these results are best interpreted as a population-level trend rather than a determination of individual halo principal axes (\citealt{2008MNRAS.385.1511W}).

\subsection{Dependence of Warps on Large-scale Filaments}
\label{sec:dep}

The triaxial, non-spherical shapes of DM halos and the anisotropic spatial distribution of satellite galaxies are significantly influenced by filamentary accretion and large-scale tidal fields (\citealt{2015MNRAS.450.2727T}; \citealt{2020MNRAS.495..502M}; \citealt{2021NatAs...5..839W}).
\citet{2013ApJ...779..160Z} found that galaxies within filaments tend to align their major axes along the filament direction, while galaxies in sheets align with the plane of the sheet, reflecting the impact of large-scale tidal forces on galaxy orientation and evolution. 
Extending this to satellite systems, \citet{2015MNRAS.450.2727T} demonstrated that satellites preferentially lie along filament axes, indicating that the cosmic web guides satellite accretion and group assembly.

Given that galactic warps are thought to arise from torques exerted by non-spherical DM halos (\citealt{1988MNRAS.234..873S}; \citealt{1989MNRAS.237..785O}; \citealt{1995MNRAS.275..897N}; \citealt{1999MNRAS.303L...7J}; \citealt{2009ApJ...696.1899J}; \citealt{2009ApJ...703.2068D}; \citealt{2010MNRAS.408..783R}; \citealt{2022MNRAS.510.1375S}; \citealt{2023MNRAS.523.5853V}), it is reasonable to suggest a continuous connection between warp structures, halo anisotropy, and the filamentary environment. 
Filaments drive anisotropic accretion and shape DM halo geometry, which in turn influences satellite distributions and disk morphology, potentially leading to warp formation. 
However, few studies have directly connected warp properties to large-scale structures. 
\citet{2008A&A...488..511L} conducted a statistical analysis of 97 edge-on galaxies from SDSS, examining the correlation between warp amplitude and galaxy inclination relative to void surfaces. 
Although they rejected the null hypothesis of no correlation at 94.4\% confidence, their results were not definitive. 
Nonetheless, their study represents an early observational effort to link cosmic void geometry---and by extension, large-scale structure---to galactic warp features, suggesting that large-scale tidal environments may influence disk warps, though further confirmation is required.

\begin{figure}
	\includegraphics[width=\columnwidth]{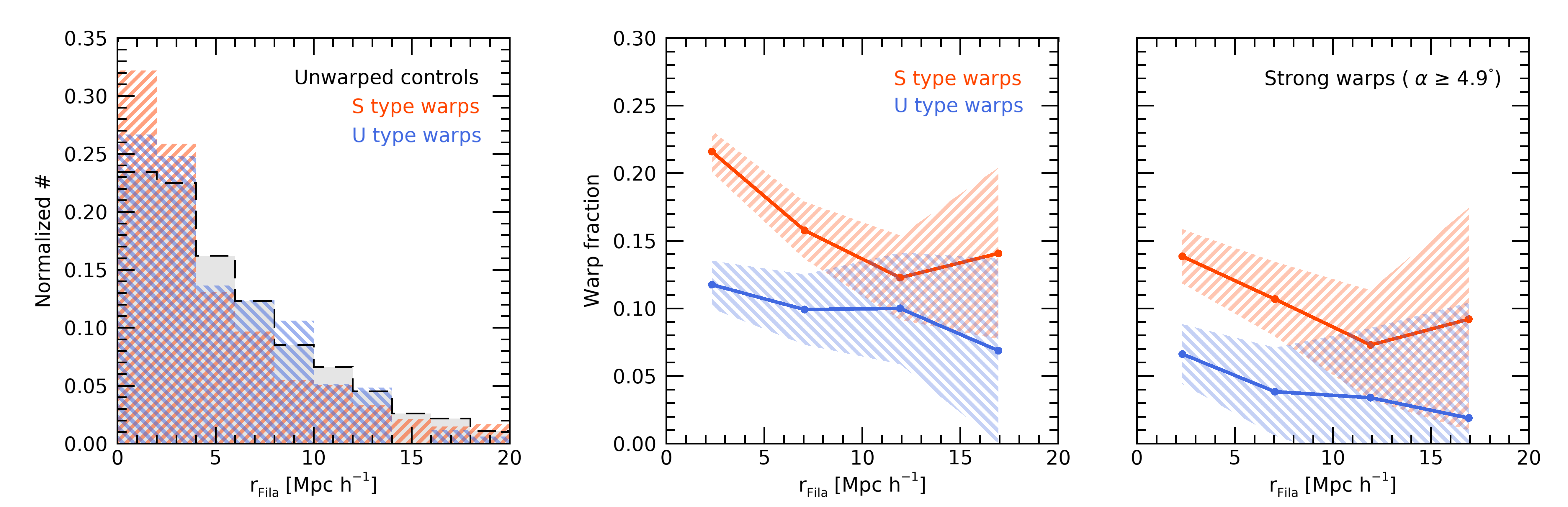}
    \caption{Left: The normalized histogram of the distance to the nearest filament, $r_\textrm{Fila}$, for the unwarped control sample (dashed black), S type (red), and U type warped galaxies (blue).
    Middle: Fraction of warped galaxies of S type (red) and U type (blue) as a function of $D_\textrm{Fila}$. 
    Shaded region represents Poisson errors in each $r_\textrm{Fila}$ bin.
    Right: Same as the middle panel, but for strongly warped galaxies ($\alpha \geq 4.9^\circ$).}
    \label{fig:4}
\end{figure}

Building on these perspectives, we investigate whether nearby filamentary structures influence the incidence of warped disk galaxies. 
Using filament identification from DisPerSE (\citealt{2020A&A...642A..19M}) applied to SDSS DR12, we measure the distance to the nearest filament, $r_{\rm Fila}$, for each central galaxy. 
Figure~\ref{fig:4} shows the relationship between filament proximity and warp occurrence. 
The left panel presents $r_{\rm Fila}$ distributions for unwarped, S and U type warped galaxies, revealing that warped galaxies are generally closer to filaments. 
There is a detectable excess of warped galaxies within $r_{\rm Fila} < 4\ \textrm{Mpc}\ h^{-1}$. 
Two-sample KS tests support this: S type versus unwarped sample yields $p\,=\,1.18 \,\times\,10^{-5}$. 
In contrast, the null hypothesis cannot be rejected for U type warps compared to the control sample ($p\,=\,0.3147$), indicating a significant preference for S type central galaxies to lie closer to filaments, with no clear evidence for U types.
The middle and right panels of Figure~\ref{fig:4} show the warp fraction as a function of $r_{\rm Fila}$, highlighting a clear trend: warp incidence increases with decreasing filament distance. 
This trend is especially pronounced for S type warps within $r_{\rm Fila} < 10 \ \textrm{Mpc} \ h^{-1}$, while U type warps exhibit no statistically significant rise.
Based on the warp measurement from \href{https://iopscience.iop.org/article/10.3847/1538-4357/ac7462/meta}{Paper I}, strongly warped galaxies with warp amplitudes above the median ($\alpha > 4.9^{\circ}$) do not show a statistically significant gradient, but their measured fractions remain consistent with a mild increase toward smaller $r_{\rm Fila}$. 
We caution, however, that given the reduced sample sizes and associated uncertainties, the apparent difference is not significant on its own.

\begin{figure}
	\includegraphics[width=\columnwidth]{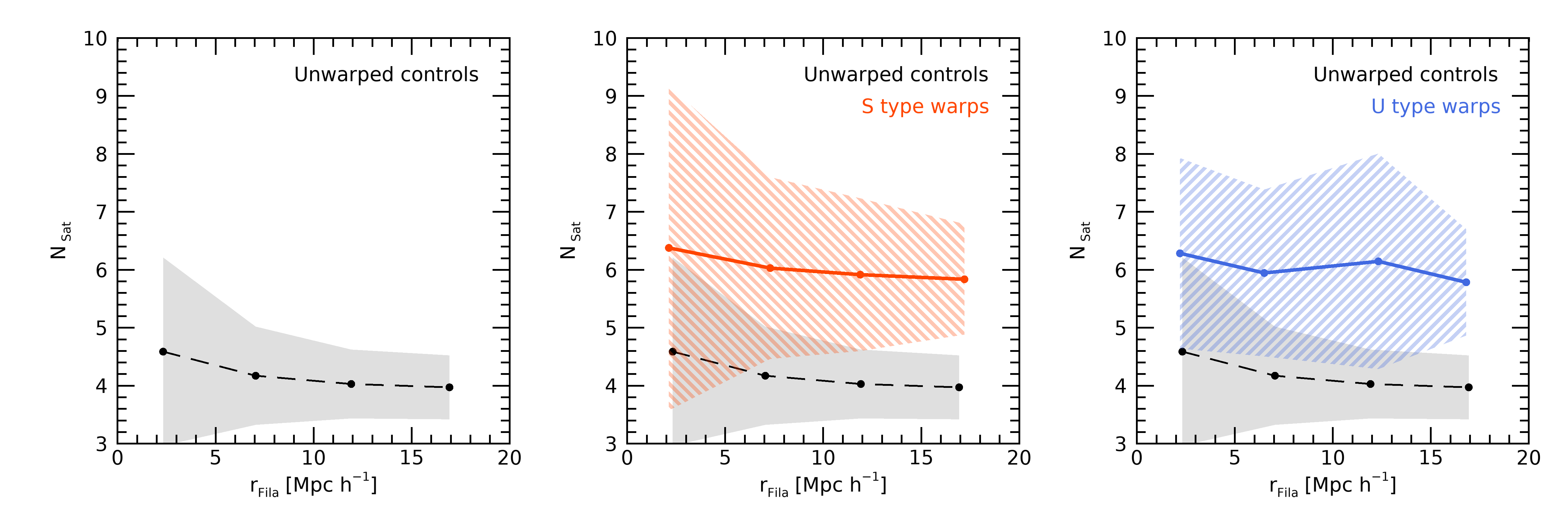}
    \caption{Left: Number of satellites per host, $N_\textrm{Sat}$, versus distance to the nearest filament, $r_\textrm{Fila}$, for the unwarped control sample (black dashed). The shaded band shows the standard deviation for each bin.
    Middle: Same as left, for S-type warped galaxies (red); the control is overplotted for comparison.
    Right: Same as left, for U-type warped galaxies (blue); the control is overplotted.}
    \label{fig:5a}
\end{figure}

The enhanced warp incidence near filaments is intriguing, but it must be tested whether this reflects genuine large-scale environmental effects or simply more frequent satellite interactions in filamentary regions. 
Galaxies close to filaments can host more satellites (\citealt{2019MNRAS.484.4325W}). 
Because this study analyzes per-host satellite geometry, and satellite numbers ($N_\textrm{Sat}$) govern both the rate of tidal encounters and the halo-shape measurements, it is necessary to examine any dependence of $N_\textrm{Sat}$ on $r_\textrm{Fila}$. 
Figure~\ref{fig:5a} shows $N_\textrm{Sat}$ versus $r_\textrm{Fila}$: the unwarped control exhibits a marginal increases toward smaller $r_\textrm{Fila}$ with large uncertainties, whereas S and U type warps show no significant trends. 
The absence of a detectable increase in $N_\textrm{Sat}$ for warped hosts is consistent with their satellites being lower-mass (see also Section~\ref{sec:excess}; Figure~\ref{fig:1}), for which a strong filament-related excess is not expected (\citealt{2019MNRAS.484.4325W}). 
Given the lack of clear $N_\textrm{Sat}$--$r_\textrm{Fila}$ trends and the density-matched construction of the unwarped control, biases from denser environments and elevated interaction rates near filaments should be minimal.
The increase in warp frequency toward smaller $r_\textrm{Fila}$ therefore likely reflects a genuine physical trend.

Examining the relative alignment between the anisotropic distribution of satellite galaxies and nearby filaments provides a direct test of the link between large-scale environments and warp evolution. 
To perform this analysis, it is essential to estimate the orientation of each edge-on central galaxy's disk accurately, as the satellite distributions were reoriented based on the central's major axis, as described above. 
In \href{https://iopscience.iop.org/article/10.3847/1538-4357/ac7462/meta}{Paper I}, we demonstrated that $PA$ of galaxies from the SDSS pipeline can be inaccurate; thus, we refined them by iteratively adjusting warp amplitudes until the derived $PA$ converged within $\Delta $PA$ \,< 0.01^{\circ}$. 
This yields a robust vector defining each galaxy's true major axis.

\begin{figure}
	\includegraphics[width=\columnwidth]{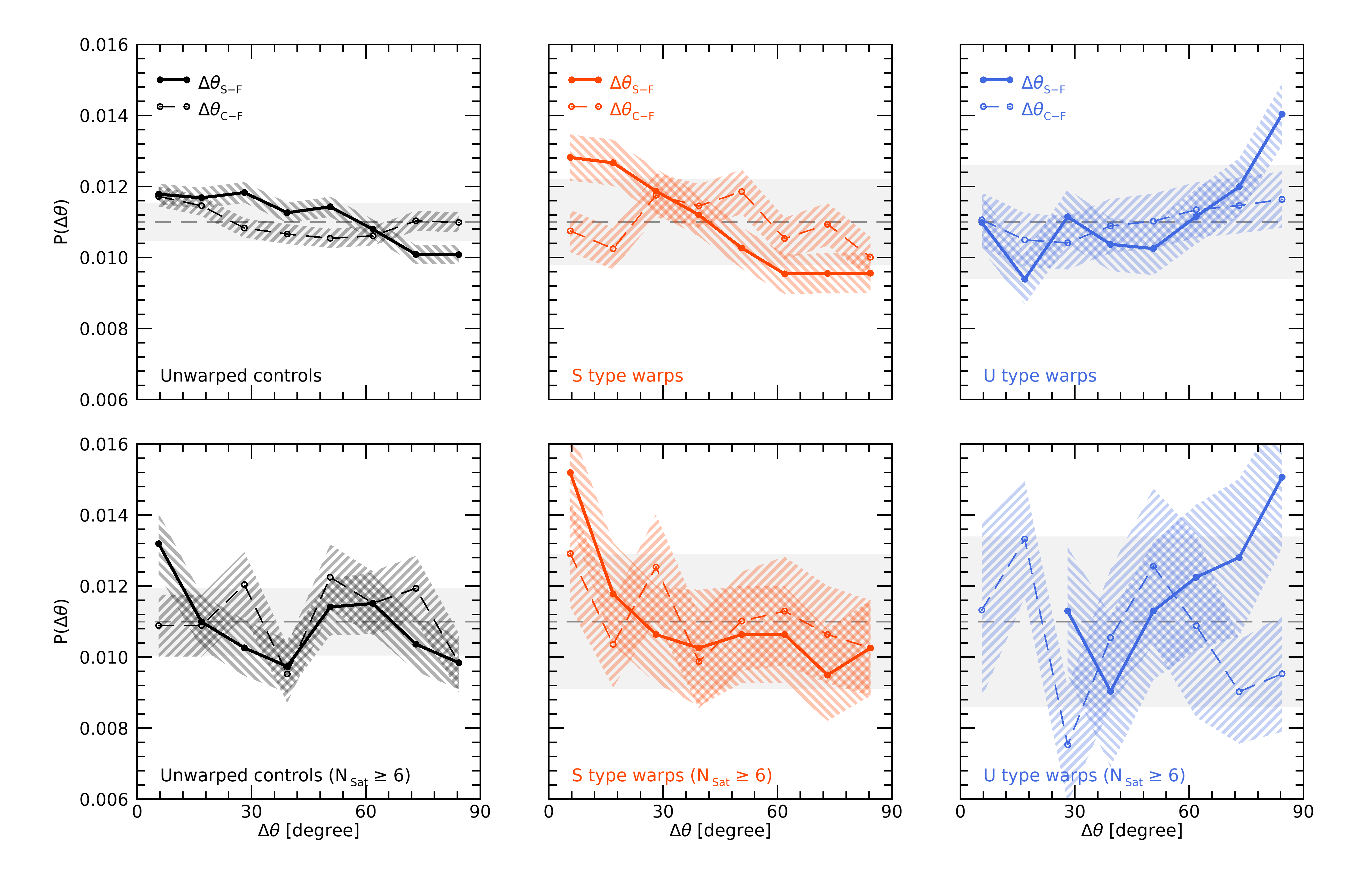}
    \caption{Upper: Probability density function, $P(\Delta \theta)$, of the misalignment angle between the mean orientation of satellite distributions ($\theta_\textrm{S-F}$, solid thick line) and the major axis of central edge-on disk galaxies ($\theta_\textrm{C-F}$, dashed line) relative to the direction of the nearest filament for the unwarped sample (black), S type (red), and U type (blue).
    Shaded region represents Poisson errors for each bin.
    For comparison, the expected isotropic distribution with the same number of central and satellite galaxies is shown as a gray horizontal line, while the $3\sigma$ range of its random distribution is shaded in gray.
    Lower: Same as upper panels, but for central galaxies that host at least six satellites ($N_\textrm{Sat}\ge 6$).}
    \label{fig:5}
\end{figure}

We define $\Delta \theta_\textrm{C-F}$ as the misalignment angle between the central galaxy's major axis and the filament $PA$ ($PA_{\rm Fila}$). 
Similarly, based on the mean direction of each satellite distribution ($\phi$), we define $\Delta \theta_\textrm{S-F}$ as the misalignment between $\phi$ and $PA_{\rm Fila}$. 
The upper panels of Figure~\ref{fig:5} present the probability distribution of misalignment angles between the major axis of central disk galaxies and nearby filaments ($\theta_\textrm{C-F}$), and between the orientation of satellite systems and filaments ($\theta_\textrm{S-F}$), for unwarped control galaxies and those with S and U type warped galaxies.
The $P(\Delta \theta_\textrm{C-F})$ distributions for unwarped galaxies show only a weak correlation, with probabilities remaining low and consistent with random orientations, suggesting that central disk alignments are largely unaffected by nearby filaments. 
This marginal trend toward $\Delta \theta_\textrm{C-F} \simeq 0^{\circ}$ is consistent with previous studies, which showed that disk galaxies generally exhibit weak or no significant alignment with large-scale structures (e.g., \citealt{2015ApJ...798...17Z}; \citealt{2016MNRAS.457..695P}; \citealt{2019ApJ...876...52K}; \citealt{2025MNRAS.538.2660B}; \citealt{2025ApJ...983..122R}; \citealt{2025arXiv250622794W}).
In contrast, $P(\Delta \theta_\textrm{S-F})$ reveals systematic trends among warped galaxies. 
For S type warps, satellites preferentially align along filaments, with a pronounced peak at $\Delta \theta_\textrm{S-F} \simeq 0^{\circ}$ reaching $P(\Delta \theta_\textrm{S-F}) \simeq 0.0128$, above the random expectation. 
U type warps show a distinct peak at $\Delta \theta_\textrm{S-F} \simeq 90^{\circ}$ reaching $P(\Delta \theta_\textrm{S-F}) \simeq 0.014$, indicating preferential perpendicular alignment.

These trends persist when we restrict the sample to systems with higher satellite counts ($N_{\rm Sat}\ge 6$), as shown in the lower panels of Figure~\ref{fig:5}, albeit with larger uncertainties owing to the reduced sample size. 
The alignment of satellite distributions around unwarped controls with nearby filaments strengthens, particularly for $\Delta\theta<45^{\circ}$. 
This is consistent with unwarped hosts tending to have more satellites adjacent to filaments (Figure~\ref{fig:5a}); imposing $N_{\rm Sat}\ge 6$ therefore biases the sample toward systems closer to filaments and amplifies the alignment signal (\citealt{2015MNRAS.450.2727T}; \citealt{2015ApJ...800..112G}; \citealt{2017A&A...597A..86P}). 
S-type warps with $N_\textrm{Sat}\ge 6$ show a pronounced rise in $P(\Delta\theta_{\rm S-F})$ toward $\Delta\theta=0^{\circ}$, reaching $P(\Delta\theta_{\rm S-F}) \simeq 0.015$, indicative of systematic alignment with the filamentary structures, while the central disks show no strong correlations. 
U-type warps still exhibit perpendicular orientations near $\Delta\theta=90^{\circ}$, with $P(\Delta\theta_{\rm S-F}) \simeq 0.015$, above the random expectation (gray bands); however, the extremely limited sample renders this result statistically inconclusive. 
In summary, the relative orientation between satellite systems and filaments varies with warp morphology, even when we restrict to rich satellite systems ($N_\textrm{Sat}\ge 6$), suggesting that satellite orientations around warped galaxies are influenced by nearby filamentary structure.

\section{Discussion: From Cosmic Web to Galactic Warp}
\label{sec:web}

\begin{figure}
	\includegraphics[width=\columnwidth]{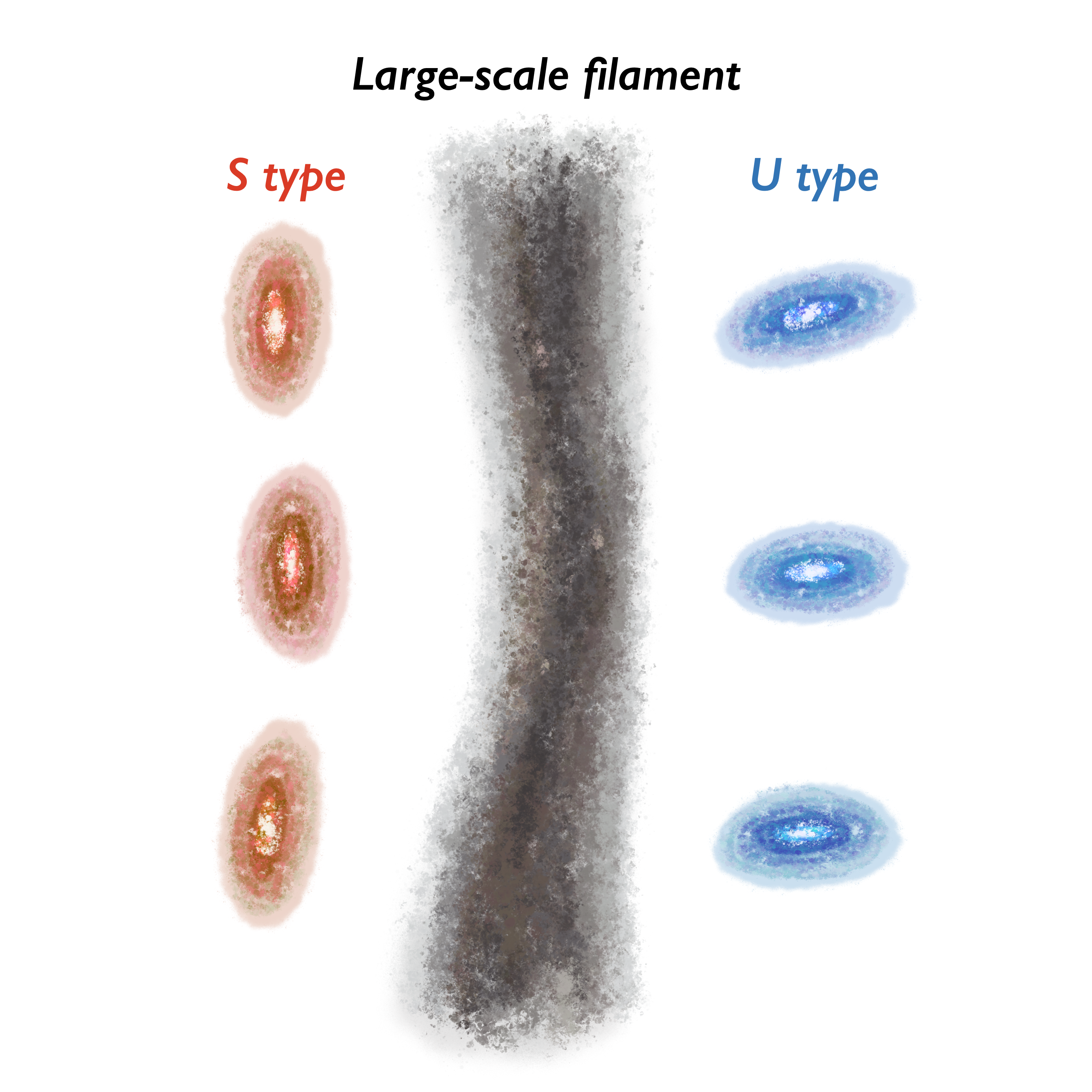}
    \caption{A schematic diagram of the relative orientation between DM halos and nearby large-scale filaments (gray shaded region) for S type (red ellipse) and U type (blue ellipse) warped galaxies. 
    The figure highlights a characteristic difference in alignment: satellites around S type warped galaxies tend to follow the direction of the nearest filament, while those around U type warped galaxies are preferentially oriented perpendicular to the filament axis.}
    \label{fig:6}
\end{figure}

Our findings reveal that satellite systems around warped disk galaxies exhibit strongly anisotropic spatial distributions relative to nearby large-scale filaments. 
This anisotropy is not uniform across warp types but instead shows a clear morphological dependence: in S type warped galaxies, satellite systems tend to align with the direction of the nearest filament, whereas in U type warped galaxies, they are preferentially oriented perpendicular to the filament spine. 
Figure~\ref{fig:6} visually depicts these distinct relative orientations of DM halos around S and U type warped galaxies, as revealed by our analysis.
While the results offer new insights into the environmental dependence of warp formation, certain limitations should be noted. 
In this study, we have inferred DM halo properties solely from the current spatial positions of satellites, without incorporating their kinematics.
Although including satellite velocities would offer a more complete picture of their dynamical state and orbital history, our primary focus was on spatial anisotropy and their relative alignment with respect to nearby large-scale structures. 
Thus, while the absence of kinematic information represents a limitation, it does not undermine the robustness of our main conclusions.
Our findings represent the first observational evidence suggesting that the relative orientation between halos and the cosmic web is systematically linked to galactic warp morphology.

In the hierarchical picture, material flows into halos along filaments (e.g., \citealt{1996Natur.380..603B}), and satellites preferentially accrete along filamentary directions aligned with host-halo major axes (\citealt{2005MNRAS.363..146L}), shaping both halo orientation and satellite planes (e.g., \citealt{2005ApJ...624..505Z}; \citealt{2013MNRAS.433..515W}). 
Our results add a morphological dependence to this framework: S type warped hosts exhibit satellite systems preferentially aligned with the local filaments, whereas U type warped hosts show satellites preferentially perpendicular to it. 
We interpret the S type case as the outcome of coherent, filamentary accretion that builds halos and satellite systems aligned with the filament—conditions that maintain relatively symmetric, long-lived warps (\citealt{2000MNRAS.311..733I}; \citealt{2004A&A...425...67R}). 
By contrast, the U type signal is consistent with more chaotic, multi-stream or transverse accretion---plausibly near nodes---where stochastic torques misaligned the disk, halo, and filament (\citealt{2010ApJ...723..364A}; \citealt{2014MNRAS.444.1453D}), decoupling halo major axes from the filament and fostering asymmetric, unstable warps (\citealt{2002A&A...386..169L}; \citealt{2003Ap&SS.284..747B}; \citealt{2009ApJ...696.1899J}; \citealt{2023MNRAS.526.5756G}).
The recently reported coherent rotation of filaments provides a physical conduit for angular-momentum delivery that can sustain the S-type configuration when inflow well aligned along the filaments (\citealt{2021NatAs...5..839W}).

The distinction between S and U type warps may reflect not only different formation mechanisms but also different stages in the evolutionary sequence of warped galaxies (e.g., \citealt{2002A&A...382..513R}; \citealt{2014ApJ...789...90K}; \href{https://iopscience.iop.org/article/10.3847/1538-4357/ac7462/meta}{Paper I}).
In this framework, U type warps are likely to form in dynamically young systems, where the DM halo has not yet settled into a stable orientation and remains misaligned or even perpendicular with the surrounding filament. 
During this early phase, recent accretion or asymmetric gravitational influences can create uneven torques on the disk, leading to asymmetric, one-sided U type warps. 
As the system evolves, the halo's orientation gradually stabilizes with respect to the surrounding filament, and the warp structure transitions into a more symmetric S type morphology (e.g., \citealt{1999ApJ...513L.107D}; \citealt{2010MNRAS.408..783R}).

This scenario is consistent with theoretical studies that predict DM halos become more aligned with nearby filaments over time (\citealt{2007MNRAS.381...41H}; \citealt{2012MNRAS.427.3320C}; \citealt{2013ApJ...762...72T}; \citealt{2021MNRAS.502.5528L}; \citealt{2021MNRAS.503.2280G}). 
For instance, \citet{2012MNRAS.427.3320C} and \citet{2015MNRAS.446.2744L} found that as structure formation proceeds, DM halos tend to become increasingly aligned with the direction of nearby filaments---especially in systems that grow gradually through smooth accretion along filamentary flows. 
In such cases, the alignment of the halo's major axis and angular momentum with the filament becomes more coherent, allowing the disk to experience more symmetric torques from its halo environment.

Thus, our observational findings can be explained as a reflection of evolutionary stage: 
U type warps form early, when the system is dynamically unsettled, and the halo is still misaligned; S type warps emerge later, after the halo has co-evolved with the filament and the system reaches a more stable torque balance.
\href{https://iopscience.iop.org/article/10.3847/1538-4357/ac7462/meta}{Paper I} found that U type warped galaxies are on average blue and less massive than S type galaxies. 
This color and mass difference supports an evolutionary pathway in which galaxies begin as low-mass, actively star-forming U type systems, and as they grow in mass and settle into their filamentary environment, they evolve into more massive, redder, and dynamically relaxed S type galaxies.

\section{Summary and Conclusion}
\label{sec:sum}

In this study, we investigate the spatial distribution of satellite galaxies around warped disk galaxies using the warped galaxy catalog established in our previous series \href{https://iopscience.iop.org/article/10.3847/1538-4357/ac7462/meta}{Paper I}, constructed from SDSS DR7.
By classifying warp morphologies into S type and U type, and employing a carefully matched control sample, we examine how satellite anisotropy correlates with warp presence and structure. 

Our key findings are as follows: 

\begin{enumerate}
\item
Warped hosts are consistent with being driven by frequent, lower-mass satellite interactions. 
They host more satellites---S type: $6.04\pm3.28$ and U type: $5.94\pm2.85$---than unwarped controls ($4.19\pm1.69$), although these averages overlap within $1\sigma$.
A two-sample KS test confirms that the distributions of satellite numbers differ from the control sample (control vs. S type: $D=0.179$, $p=4.6\times10^{-14}$; control vs. U type: $D=0.231$, $p=1.24\times10^{-10}$).
Their satellites are also systematically less massive---S type: $\langle \log (M_{\rm sat}/M_{\rm cent}) \rangle = -1.32\pm0.45$ and U type---compared to unwarped counterparts, which average $\langle \log (M_{\rm sat}/M_{\rm cent}) \rangle = -0.78\pm0.55$.
Likewise, two-sample KS tests show significant difference in the mass ratio distributions (control vs. S type: $D=0.233$, $p=1.7\times10^{-21}$; control vs. U type: $D=0.181$, $p=7.3\times10^{-10}$).

\item 
Stacked satellite distributions around warped galaxies suggest anisotropic features, whereas unwarped controls do not show clear deviations from isotropy within the current uncertainties.
The probability density $P(\phi)$ of satellite position angles relative to the host major axis is nearly flat (deviation $\sigma \approx 1.42\times10^{-5}$) for unwarped systems. 
In contrast, both S and U type hosts show an excess at $45^{\circ}<\phi<90^{\circ}$, reaching $P(\phi)\simeq 0.003$, supporting a role for anisotropic halo potentials and associated tidal effects.
We caution, however, that these differences weaken when restricting to $N_{\rm Sat}\ge 6$; thus, they should not be interpreted as strong trends until more observational data become available.

\item 
Warps occur more frequently---particularly S types---in galaxies closer to large-scale filaments ($r_{\rm Fila} < 4\ \textrm{Mpc}\ h^{-1}$). 
Although unwarped controls marginally host more satellites adjacent to filaments, warped galaxies do not show a systematic satellite–filament proximity bias with morphology. 
Thus, the enhanced warp incidence near filaments is not an artifact of higher satellite abundance there, but a genuine effect.

\item 
The alignment of satellite systems with nearby filaments, $\Delta\theta_{\rm S-F}$, depends on warp morphology. 
Unwarped hosts show only weak or no filament-related signal. 
By contrast, S type warps preferentially align along filaments, peaking at $\Delta\theta_{\rm S-F}=0^\circ$ with $P(\Delta\theta_{\rm S-F})\simeq0.0128$, whereas U type warps favor perpendicular alignment, peaking at $\Delta\theta_{\rm S-F}=90^\circ$ with $P(\Delta\theta_{\rm S-F})\simeq0.014$; both exceed the random expectation. 
These trends persist in richer systems ($N_\textrm{Sat}\geq 6$), albeit with larger uncertainties, indicating a genuine large-scale effect; a larger sample will provide stronger statistics to test this trend.

\end{enumerate}

Our results suggest that cosmologically anisotropic satellite distributions, plausibly connected to nearby filaments, provide an additional pathway for warp formation, complementing the ubiquity of warps not explained by galaxy–galaxy interactions alone.
The stacked satellite distributions reveal a clear anisotropy around warped hosts (in contrast to unwarped controls), directly indicating that warped galaxies preferentially inhabit lopsided---oblate or misaligned---halo configurations long proposed in the literature as the origin of warps and, at least in part, as mechanisms that sustain warp structures over long timescales.
Although such halo-driven mechanisms have been difficult to test observationally, our measurements supply a plausible framework.
Moreover, the fact that the orientation of the satellite anisotropy aligns or is perpendicular to nearby filamentary structures argues against a stochastic, galaxy-by-galaxy origin and instead points to regulation by the cosmic web.
To explain the observed anisotropic distribution of satellites and the differences in their relative orientation around S and U type warped galaxies with respect to nearby large-scale filaments, we consider several scenarios, including anisotropic tidal shaping and time evolution of warps.
These probabilities provide a cohesive link that connects the global to local evolution of galaxies, connecting the morphology and orientation of galactic warps to their halo configurations and the cosmic web.

In future work, larger samples of edge-on disk galaxies extending to higher redshifts (e.g., \citealt{2017ApJ...846..159B}; \citealt{2024arXiv241103430F}; \citealt{2025MNRAS.540.3493T}), in collaboration with our extensive citizen-science classification project (\textit{Poppin' Galaxy}), along with spectroscopy and kinematics of individual satellites, and high-resolution cosmological simulations (e.g., Illustris TNG; \citealt{2019MNRAS.490.3234N}) will permit more tests of our conclusions.
Upcoming deep image surveys, such as Euclid and Dark Energy Spectroscopic Instrument, will enable more robust statistical analysis of warped galaxies by providing deeper observations and improved redshift information. 
Together, these datasets will trace how filamentary accretion shapes halos and satellite systems and how warp structures evolve over cosmic time.

\begin{acknowledgments}
This work was supported by the faculty research fund of Sejong University in 2025.
SLJ acknowledges the support of a UKRI Frontiers Research Grant [EP/X026639/1], which was selected by the European Research Council, and the STFC consolidated grants [ST/S000488/1] and [ST/W000903/1].
This work makes use of the service L3S/COSFIL, the Large Scale Structure Services developed by OSUPS, OCA, Pytheas and ByoPiC products.
S.-J.Y. and S.P. acknowledge support from the Mid-career Researcher Program (RS-2024-00344283 and RS-2023-00208957, respectively) through Korea's NRF funded by the Ministry of Science and ICT and support from the Basic Science Research Program (RS-2022-NR070872) through Korea's NRF funded by the Ministry of Education. 
Funding for the Sloan Digital Sky Survey (SDSS) has been provided by the Alfred P. Sloan Foundation, the Participating Institutions, the National Aeronautics and Space Administration, the National Science Foundation, the U.S. Department of Energy, the Japanese Monbukagakusho, and the Max Planck Society.
The SDSS Web site is \url{http://www.sdss.org/}.
The SDSS is managed by the Astrophysical Research Consortium (ARC) for the Participating Institutions. 
The Participating Institutions are The University of Chicago, Fermilab, the Institute for Advanced Study, the Japan Participation Group, The Johns Hopkins University, Los Alamos National Laboratory, the Max-Planck-Institute for Astronomy (MPIA), the Max-Planck-Institute for Astrophysics (MPA), New Mexico State University, the University of Pittsburgh, Princeton University, the United States Naval Observatory, and the University of Washington.
The data in this paper are the result of the efforts of the Galaxy Zoo 2 volunteers, without whom none of this work would be possible. Their efforts are individually acknowledged at \url{http://authors.galaxyzoo.org}.
This work makes use of the service L3S/COSFIL, the Large Scale Structure Services developed by OSUPS, OCA, Pytheas, and ByoPiC products.
We sincerely thank the talented graphic designer, Zina Kim, spouse of the first author, for her outstanding contribution to the visual representation in Figure~\ref{fig:6}.
\end{acknowledgments}

\end{document}